# COV-ECGNET: COVID-19 detection using ECG trace images with deep convolutional neural network.

Tawsifur Rahman[1], Alex Akinbi[2], Muhammad E. H. Chowdhury[1*], Tarik A. Rashid[3], Abdulkadir Şengür[4], Amith Khandakar[1], Khandaker Reajul Islam[1], Aras M. Ismael[5*]

[1]Department of Electrical Engineering, Qatar University, Doha-2713, Qatar
[2]School of Computer Science and Mathematics, Liverpool John Moores University, United Kingdom
[3]Computer Science and Engineering Department, School of Science and Engineering, University of Kurdistan Hewler, Erbil, KRG, Iraq.
[4]Firat University, Technology Faculty, Electrical-Electronics Engineering Department, Elazig, Turkey
[5]Sulaimani Polytechnic University, College of Informatics, Information Technology Department, Sulaymaniyah, Iraq

**\***Correspondence: Aras M. Ismael (aras.masoud@usa.com);
                Muhammad E. H. Chowdhury (mchowdhury@qu.edu.qa)

## Abstract

The reliable and rapid identification of the COVID-19 has become crucial to prevent the rapid spread of the disease, ease lockdown restrictions and reduce pressure on public health infrastructures. Recently, several methods and techniques have been proposed to detect the SARS-CoV-2 virus using different images and data. However, this is the first study that will explore the possibility of using deep convolutional neural network (CNN) models to detect COVID-19 from electrocardiogram (ECG) trace images. In this work, COVID-19 and other cardiovascular diseases (CVDs) were detected using deep-learning techniques. A public dataset of ECG images consists of 1937 images from five distinct categories, such as Normal, COVID-19, myocardial infarction (MI), abnormal heartbeat (AHB), and recovered myocardial infarction (RMI) were used in this study. Six different deep CNN models (ResNet18, ResNet50, ResNet101, InceptionV3, DenseNet201, and MobileNetv2) were used to investigate three different classification schemes: i) two-class classification (Normal vs COVID-19); ii) three-class classification (Normal, COVID-19, and Other CVDs), and finally, iii) five-class classification (Normal, COVID-19, MI, AHB, and RMI). For two-class and three-class classification, Densenet201 outperforms other networks with an accuracy of 99.1%, and 97.36%, respectively; while for the five-class classification, InceptionV3 outperforms others with an accuracy of 97.83%. ScoreCAM visualization confirms that the networks are learning from the relevant area of the trace images. Since the proposed method uses ECG trace images which can be captured by smartphones and are readily available facilities in low-resources countries, this study will help in faster computer-aided diagnosis of COVID-19 and other cardiac abnormalities.

## 1. Introduction

Coronavirus Disease 2019 (COVID-19) has rapidly spread with increased fatalities across the world leading to a long-lasting global pandemic. Over 166 million cases have been recorded as of May 21, 2021, with over 3.4 million fatalities documented worldwide [1]. The Severe acute respiratory syndrome coronavirus 2 (SARS-CoV-2) virus mostly affects the respiratory system, but it can also lead to multi-organ failure. It has a severe impact on the cardiovascular system [2-6]. The advancement of artificial intelligence in biomedical applications has helped in developing trained networks for reliable computer-aided diagnostic decisions and thus reducing the pressure from the healthcare facilities (such as medical doctors, healthcare staff, etc.) [7]. Several deep learning models have been proposed in recent studies to identify abnormalities from medical images, including chest X-ray images and computerized tomography (CT) scans [8]. Degerli et al. in [9], introduced a novel approach for the combined localization, severity grading, and detection of COVID-19 from 15495 CXR pictures by constructing so-called infection maps, which can accurately localize and grade the severity of COVID-19 infection with 98.69 percent accuracy. For chest X-ray image classification, Kesim et al. proposed a novel convolutional neural network (CNN) model [10]. Since pre-trained CNN models have difficulty in practical applications, the authors designed a small-sized CNN architecture that showed very promising performance in classifying twelve different abnormalities from chest X-ray images (Atelectasis, Cardiomegaly, Consolidation, Edema, Effusion, Emphysema, Fibrosis, Infiltration, Mass, Nodule, Pleural Thickening, Pneumothorax) and reported an accuracy score of 86 %. Liu et al. proposed a tuberculosis (TB) detection technique using chest X-ray and deep learning models[11]. The authors proposed a new CNN model and utilized shuffle sampling to deal with the imbalanced dataset issue and yielded an accuracy score of 85.68 %. Rahman et al.[12] applied various pre-trained CNNs to categorize CXR pictures as having pulmonary tuberculosis (TB) symptoms or as being healthy. A dataset of 3,500 infected and 3,500 normal CXR pictures were used to train the suggested model. DenseNet201, the best-performing model, achieved a high detection performance of 98.57 percent sensitivity and 98.56 percent specificity. Chowdhury et al. [13] have created a public dataset consisting of Normal, Viral Pneumonia and COVID-19 chest X-ray images and used deep CNN models for binary and three class classifications. On the created dataset, transfer learning using pre-trained Squeezenet, Mobilenetv2, Inceptionv3, Chexnet, ResNet, and Densenet201 models were examined. While binary classification had an accuracy score of 99.7 %, three-class classification tasks showed an accuracy of 97.9%. Xu et al. in [14] devised a method for detecting abnormalities in the chest X-ray images. To avoid the over-fitting problem in transfer learning, the authors suggested a hierarchical-CNN model called CXNet-m1. The proposed CNN models were shallower than the pre-trained CNN models. Moreover, a novel loss function and CNN kernel improvement were introduced with an overall accuracy of 67.6 %. In the study by Rahman et al. [15], the authors reported three schemes of classifications: normal vs. pneumonia, bacterial vs. viral pneumonia, and normal, bacterial, and viral pneumonia. Normal and pneumonia images, bacterial and viral pneumonia images, and normal, bacterial, and viral pneumonia images had classification accuracy of 98 %, 95%, and 93.3%, respectively. Chouhan et al. [16] used deep learning models to detect pneumonia in chest X-ray images using five deep transfer learning models and their ensemble. The accuracy of the ensemble deep learning model was 96.4 %. Rajpurkar et al. created ChexNet, a 121-layer CNN architecture for stratifying fourteen distinct lung diseases using chest X-ray images [17]. The authors used the chest X-ray dataset to train the 121-layered DenseNet-121 CNN model. This is the first pre-trained ImageNet model which has been made public re-trained on chest X-ray images. The proposed model produces an area under the curve (AUC) values ranging from 0.704 to 0.944. Li et al. proposed a multi-resolution CNN (MR-CNN) for lung nodule identification [18]. To extract the features, a patch-based MR-CNN model was utilized, and multiple fusion approaches were applied for classification. Free Response Receiver Operating Characteristics (FROC) curve was used for performance evaluation with AUC and Refined Competition Performance Metric (R-CPM) measures of 0.982 and 0.987, respectively.

Bhandary et al. tweaked the AlexNet model to detect lung anomalies using chest X-ray images [19]. A new threshold filter and feature ensemble technique were deployed to achieve a classification accuracy of 96%. Ucar et al. [20] employed Laplacian Gaussian filters to improve the classification performance of the CNN models in chest X-ray image classification, which achieved a classification accuracy of 82.43%. Ismael and Şengür [21] demonstrated different deep learning approaches to detect COVID-19 from chest X-ray images using a Kaggle dataset and obtained the highest accuracy score 92.63%, which was produced by the ResNet50 model. Their COVID-19 detection was carried out using a variety of multiresolution techniques (Contourlet transform and Wavelet and Shearlet). Extreme Learning Machine (ELM) was applied in the classification stage and the experiment results showed that Wavelet and Shearlet can obtain higher accuracy of 92%.

COVID-19 infection can cause acute myocarditis in apparently healthy people [22]. Up to 27.8% of COVID-19 patients had an increased troponin level beyond the 99$^{th}$ percentile of the upper reference limit, indicating acute myocardial injury in an early case reported from China [23,24]. This is about ten times greater than the influenza rate (2.9%) [25]. Most COVID-19 patients, even those who have biochemical evidence of acute myocardial damage, have a moderate illness history and recover without overt cardiac problems. It is unclear if COVID-19 survivors with no overt cardiac signs have any subclinical or hidden cardiac injury that might impair long-term results. As the pandemic slows, it is critical to figure out if cardiac monitoring in COVID-19 survivors is necessary or not. The potential to screen the general population and give an extra opinion for health care practitioners is the benefit of automated 12-lead electrocardiogram (ECG) diagnostic techniques. Since 1957, attempts have been made to automate the interpretation of ECG recordings, with a focus on findings links to atrial fibrillation (AF). However, the performance of currently available automated methods has been mediocre [26]. Subsequently, despite current technological advancements, notably in the fields of sophisticated machine learning and artificial intelligence (AI) methodologies, the clinical value of automated ECG interpretations remains limited [27,28], and cardiologists continue to analyze and interpret 12-lead ECG recordings using traditional methods. In a recent work by Du et al. in [29], the approach of deep learning on ECG trace images have been explored with promising result. This work will further explore the possibility of COVID-19 and other cardiac abnormality detection with the help of deep learning techniques.

The remainder of this paper is organized as follows: Section 2 discusses the material and methods of the study, while Section 3 outlined the experimental pipeline and evaluation metrics. Section 4 discusses the results and finally, Section 5 concludes the paper.

## 1.1. Deep convolutional neural networks-based transfer learning

Six popular deep learning pre-trained CNN models have been used for COVID-19 detection using ECG trace images. These are ResNet18, ResNet50, ResNet101 [6], DenseNet201 [30], InceptionV3 [31], and MobileNetV2 [32], which are initially trained on ImageNet database. The Residual Network (also known as ResNet) was created to overcome the vanishing gradient and degradation problem [6]. ResNet has different variants based on the number of layers in the residual network: ResNet18, ResNet50, ResNet101, and ResNet152. ResNet is widely utilized for transfer learning in biomedical image classification. During training, deep neural network layers typically learn low or high-level features, whereas ResNet learns residuals rather than features [22]. Figure 1 shows the architecture of a convolutional neural network.

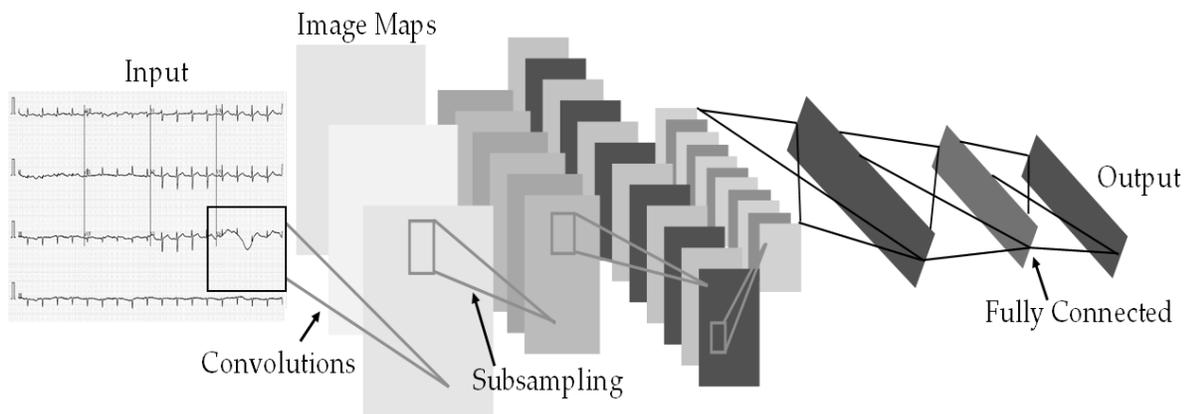

Figure 1: Architecture of a Convolutional Neural Network.

A Dense Convolutional Network (or DenseNet) [30] requires fewer parameters than a traditional CNN as this does not require training on redundant feature maps. The DenseNet has very narrow layers, hence it only adds a small number of new feature maps. DenseNet has four different known variants: DenseNet264, DenseNet169, DenseNet121 and DenseNet20. DenseNet provides straight access to the original input image as well as gradients from the loss function in each layer. As a result, the computational cost of DenseNet has been significantly lowered, making it a superior choice for image classification.

Alternatively, MobileNetv2 [32] is not comparable to other networks in-depth, rather this is a compact network. Except for the first layer, which is a full convolution, the rest of the layers are non-convolutional. Except for the last fully connected layer, which has no nonlinearity and feeds into a Softmax layer for classification, the MobileNet structure is constructed on depth-wise separable convolutions. Batch normalization and Rectified Linear Units (ReLU) nonlinearity are applied to all layers. Before the fully connected layer, a final average pooling reduces the spatial resolution to 1. MobileNet has 28 layers when depth-wise and pointwise convolutions are counted separately. Inception Networks use inception blocks to allow for deeper networks and more efficient computation by reducing dimensionality with layered convolutions.

## 1.2. Visualization Techniques

There is an increased interest in the internal mechanics of CNNs and the rationale for the models' judgments for classification. The visualization techniques aid the interpretation of CNN decision-making processes by providing a more visual representation. These also improve the model's transparency by presenting the reason behind the inference in a way that is easily understandable by human, hence enhancing confidence in the neural network's conclusions. SmoothGrad [11], Grad-CAM [12], Grad-CAM++ [13], and Score-CAM [25] are examples of visualization approaches. Because of its promising performance, Score-CAM was used in this study. The outcome is formed by a linear combination of weights and activation maps, with each activation map's weight determined by its forward passing score on the target class and eliminates the dependency of gradients. A sample image visualization with Score-CAM is shown in Figure 2, where the heat map indicates that the region dominantly contributed to the decision making in CNN. This can be useful for understanding how the network makes decisions and for enhancing end-user confidence when it can be confirmed that the network makes decisions using the important segment of ECG trace image all the time.

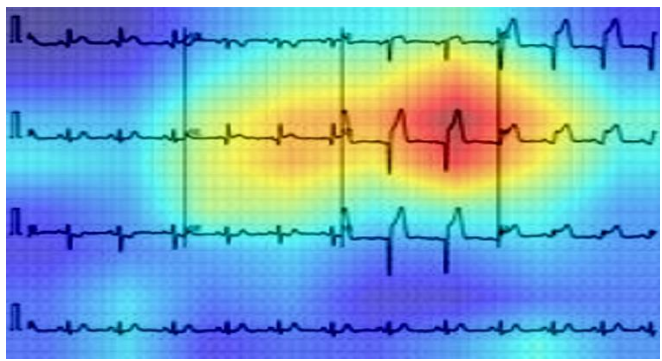

Figure 2: Score-CAM heat map on ECG trace images to show the important region for making the decision by the CNN.

## 1.3. Different abnormalities in ECG images

In this study, five distinct types of ECG trace images were used in this study, where four out of five are abnormal (COVID-19, myocardial infarction, abnormal heartbeat, and recovered myocardial infarction) and the other one is normal ECG trace images. In clinical terms, a normal ECG trace image represents the ECG of the normal person, who has no abnormality in the ECG trace. Myocardial Infarction (MI), often known as a heart attack, is a form of acute coronary syndrome that defines a sudden or short-term reduction or disruption of blood flow to the heart, causing significant damage to the heart and can be detected by ECG sensing for correct patient diagnosis [33]. Chest pain or discomfort is the most prevalent symptom, which might spread to the shoulder, arm, back, neck, or jaw. Other than the MI ECG trace images of individuals, the dataset includes ECG traces images of the patients who have just recovered from COVID-19 and are experiencing symptoms of shortness of breath or respiratory sickness and the patients who are suffering from other abnormal heartbeats. Moreover, ECG trace images of the patients who are recently recovered from myocardial infarction were also available.

Most types of cardiac abnormalities have slight variances in ECG signals, nonetheless, these tiny distinctions (e.g., a peak-peak interval or a particular wave) are frequently used for defining the variables in abnormalities classification, such as ST-segmentation change, P wave height, and T wave abnormality. Figure 3 shows two examples of aberrant kinds that may be recognized by important components. Due to its inability to efficiently gather important and discriminative aspects, deep CNN models' effectiveness is restricted when dealing with picture data challenges.

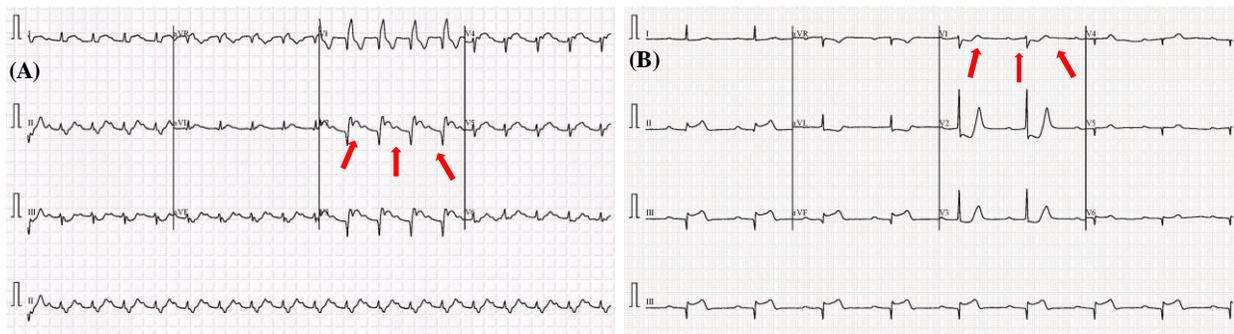

Figure 3: Illustration of two examples of ECG classes: abnormal ECG trace image for the (a) COVID-19 and (b) Myocardial Infarction patients. The subtle signs identified (highlighted in red and pointing via arrows) as key parts to detect the abnormalities.

## 2. Methodology

Figure 4 summarizes the methodology of this study. As explained earlier, this work has presented three different experimental schemes: i) binary or two-class classification (Normal vs COVID-19); ii) three-class classification (Normal, COVID-19, and Cardiac abnormalities) and finally, iii) five-class classification (Normal, COVID-19, MI, AHB, and RMI). Six state-of-the-art CNN models were trained, validated, and tested to detect abnormality from ECG trace images for each of the classification schemes.

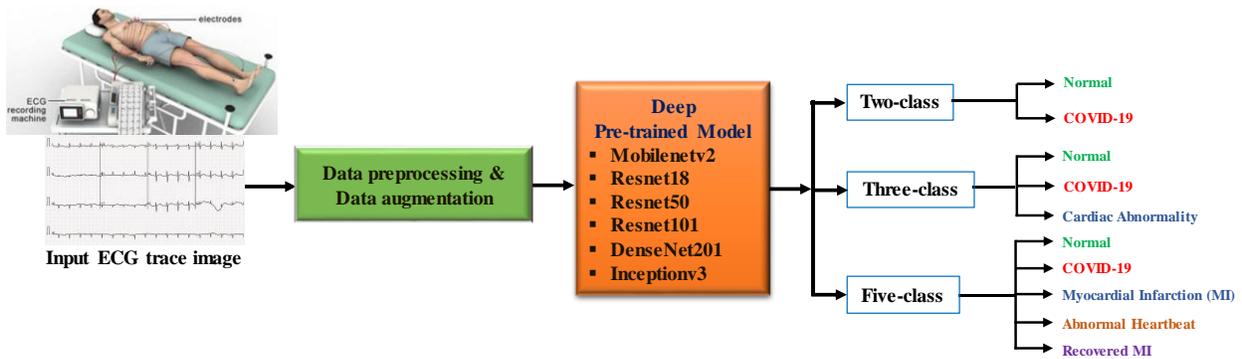

Figure 4: Overview of the methodology.

The methodology of this research work is described in the following subsections.

### 2.1. Dataset description

In this study, an ECG image dataset [34] of cardiac and COVID-19 patients is used, which consists of 1937 distinct patient records with five distinct categories (Normal, COVID-19, myocardial infarction (MI), abnormal heartbeat (AHB), and recovered myocardial infarction (RMI). All the data were collected using the ECG device 'EDAN SERIES-3' installed in Cardiac Care and Isolation Units of different health care institutes across Pakistan. Twelve lead ECG trace images were collected and were manually reviewed by medical professors using a telehealth ECG diagnostic system, under the supervision of senior medical professionals with experience in ECG interpretation. Table 1 shows the number of images for different categories in the dataset and some sample images is shown in Figure 5.

Table 1: Dataset description

| Category | Number of images | Sample rate | Leads |
|---|---|---|---|
| Normal | 859 | 500 Hz | 12 leads |
| COVID-19 | 250 | | |

| | | | | |
|---|---|---|---|---|
| Myocardial infarction | 77 | | | |
| Abnormal Heartbeat | 548 | | | |
| Recovered MI | 203 | | | |

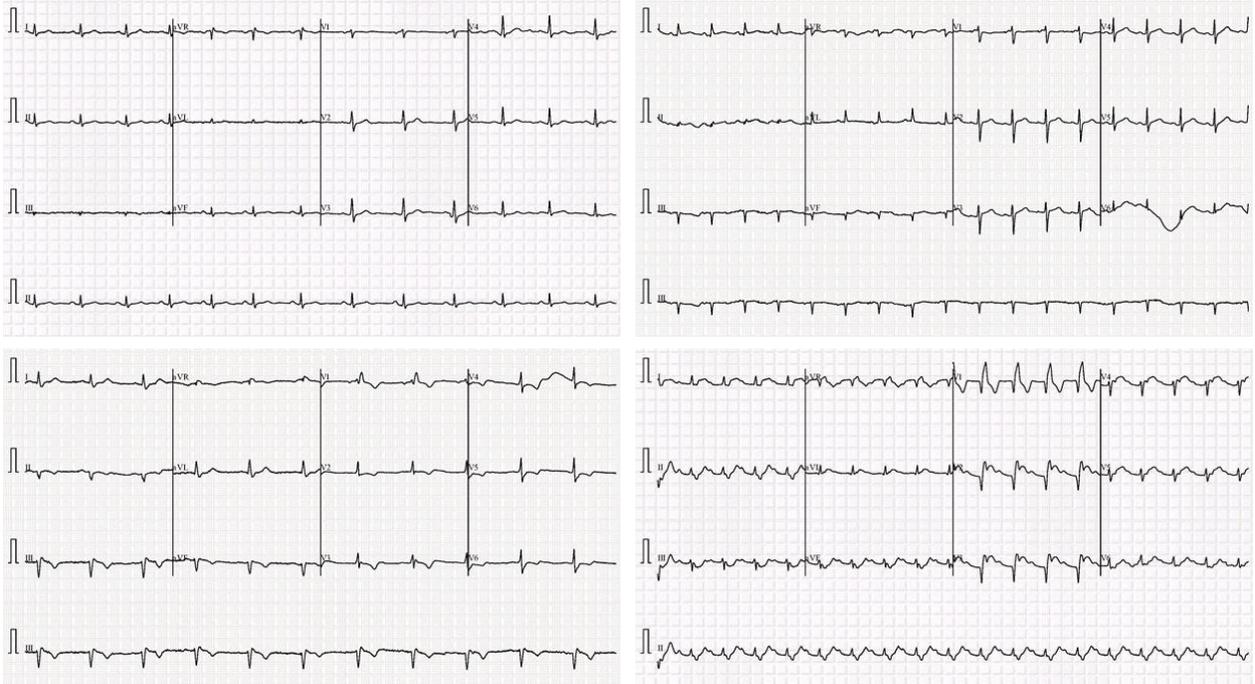

Figure. 5. Sample ECG trace images from the dataset. The horizontal axis represents time, and each time step is represented by a vertical line that lasts 0.04 seconds. Signal magnitudes in millivolts (mV), are represented on the vertical axis.

## 2.2. Preprocessing

To improve the ECG image quality, the files are preprocessed using a gamma correction enhancement technique [35]. In image normalization, linear operations, individual pixels are frequently subjected to operations such as scalar multiplication, addition, and subtraction. Gamma correction is a non-linear procedure that is applied to the pixels of a source image. To improve the image, gamma correction employs the projection relationship between the pixel value and the gamma value according to the internal map, as shown in Figure 6. If A represents the pixel value within a range of 0-255, which represents an angle value. If X represents the grayscale value of the pixel (A), then Equation (1-5) is correct. Let $X_m$ be the midpoint of the range [0, 255]. P is the linear map from group consists of the following elements:

$$\varphi: A \to \Omega, \Omega = \{\omega | \omega = \varphi(x)\}, \varphi(x) = \frac{\pi x}{2X_m} \quad (1)$$

The mapping from $\Omega$ to $\Gamma$ is defined as:

$$h: \Omega \to \Gamma, \Gamma = \{\gamma | \gamma = h(X)\} \quad (2)$$

$$\begin{cases} h(X) = 1 + f_1(X) & (3) \\ f_1(X) = \mathbf{acos}(\varphi(X)) & (4) \end{cases}$$

Based on this map, group A can be related to $\Gamma$ group pixel values. The arbitrary pixel value is calculated with a given Gamma number. Let $\gamma(X) = h(X)$, and the Gamma correction function is as follows:

$$s(X) = 255 \left(\frac{X}{255}\right)^{1/\gamma(X)} \quad (5)$$

Where $s(x)$ represents the output pixel correction value in grayscale. After gamma correction, the dataset is processed to resize the ECG images to fit the input image-size requirements of CNN networks (e.g., 224 by 224 for residual and dense networks, and 299 by 299 for inception network). Using the mean and standard deviation of the images, Z-score normalization of the image was carried out [36].

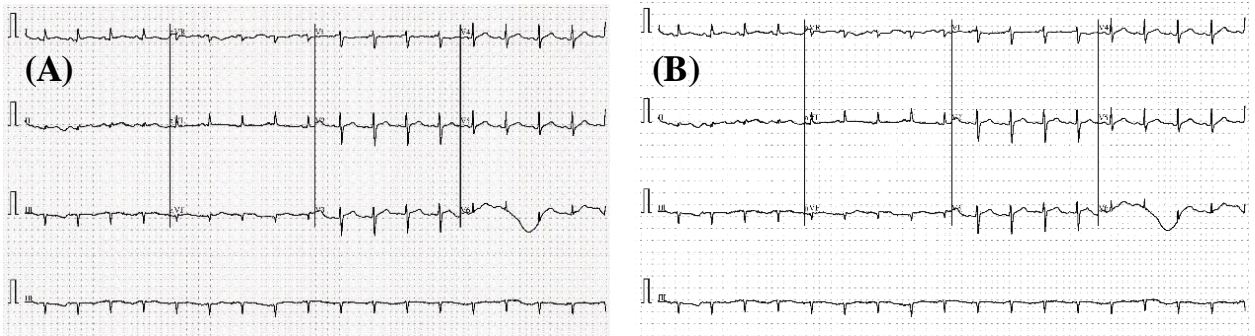

Figure 6: Preprocessing the input image: original ECG trace image (A) and Gamma corrected image (B).

## 2.3. Augmentation

Since the dataset is not balanced and the dataset does not have a similar number of images for the different categories, training with an imbalanced dataset can produce a biased model. Thus, data augmentation for the training set can help in having a similar number of images in the various classes, which can provide reliable results as stated in many recent publications [12,13,15,35,37,38]. In this study, three augmentation strategies (rotation, scaling, and translation) were utilized to balance the training images. The rotation operation used for image augmentation was done by rotating the images in the clockwise and counterclockwise direction with an angle between 5 to 10 degrees. The scaling operation is the magnification or reduction of the frame size of the image and 2.5% to 10% image magnifications were used in this work. Image translation was done by translating images horizontally and vertically by 5% to 20%.

## 3. Experiments

As discussed in Section 2, three different classification schemes were carried out in this study: two classes (normal vs COVID-19), three classes (normal, COVID-19, and cardiac abnormality), and five classes (normal, COVID-19, myocardial infarction, abnormal heartbeat, and recovered myocardial infarction) classification using different deep learning algorithms. Five-fold cross-validation was used and therefore, 80% of data were used for training and 20 % for testing. Out of the training dataset subset, 10% were utilized for validation to avoiding overfitting issues [39]. Finally, the results were a weighted average of five folds. Table 2 shows the details of the number of training, validation, and test ECG images used.

Table 2: Details of training, validation, and test set for different classification problem

| Classification | Types | Total No. of images/ class | Train set count/fold | Validation set count/fold | Test set count/ fold |
|---|---|---|---|---|---|
| Two-class | Normal | 859 | 619*4=2476 | 68 | 172 |
| | COVID-19 | 250 | 180*14=2520 | 20 | 50 |
| Three-class | Normal | 859 | 619*4=2476 | 68 | 172 |
| | COVID-19 | 250 | 180*14=2520 | 20 | 50 |
| | Abnormal | 828 | 597*4=2388 | 66 | 165 |
| Five-class | Normal | 859 | 619*4=2476 | 68 | 172 |

| | COVID-19 | 250 | 180*14=2520 | 20 | 50 |
| | Myocardial Infarction | 77 | 56*43=2408 | 6 | 15 |
| | Abnormal HB | 548 | 395*6=2370 | 44 | 109 |
| | Recovered MI | 203 | 147*17=2499 | 16 | 40 |

The networks were built with the PyTorch library and Python 3.7 on an Intel® Xeon® CPU E5-2697 v4 @ 2,30GHz with 64 GB RAM and a 16 GB NVIDIA GeForce GTX 1080 GPU. All networks were trained using the Adam optimizer with a learning rate of $10^{-3}$, a dropout rate of 0.2, a momentum update of 0.9, a mini-batch size of 16 images with 15 backpropagation epochs, and an early stopping threshold of 8 maximum epochs when no improvement in validation loss was seen. Table 3 summarizes the training settings used in the categorization studies.

Table 3: Summary of training parameters for classification experiments

| Training parameter | Learning rate | Batch size | Epochs | Epoch patience | Stopping criteria | Optimizer |
|---|---|---|---|---|---|---|
| | 0.001 | 16 | 15 | 8 | 8 | ADAM |

## 3.1. Performance Matrices for Classification

In this study, six CNN models were trained and assessed using five-fold cross-validation. After the training phase, the performance of multiple networks for the testing dataset was assessed and compared using six performance indicators, such as accuracy, sensitivity or recall, specificity, precision (PPV), and F1 score. Equations (6-10) [35] indicate the different matrices for performance evaluation:

$$Accuracy = \frac{TP + TN}{(TP + FN) + (FP + TN)} \quad (6)$$

$$Sensitivity = \frac{(TP)}{(TP + FN)} \quad (7)$$

$$Specificity = \frac{(TN)}{(TN + FP)} \quad (8)$$

$$Precision = \frac{(TP)}{(TP + FP)} \quad (9)$$

$$F1\_score = \frac{(2 * TP)}{(2 * TP + FN + FP)} \quad (10)$$

Here, for two-class, true positive (TP) is the number of correctly classified COVID-19 ECG images and true negative (TN) is the number of correctly classified normal images. False-positive (FP) and false-negative (FN) are the misclassified normal and COVID-19 ECG images, respectively. For the three-class, true positive (TP) is the number of correctly classified COVID-19 ECG images and true negative (TN) is the number of correctly classified other two classes (normal and abnormal images). False-positive (FP) and false-negative (FN) are the misclassified other two classes (normal and abnormal images) and COVID-19 ECG images, respectively. For the five-class, true positive (TP) is the number of correctly classified COVID-19 ECG images and true negative (TN) is the number of correctly classified other four classes (normal, myocardial infarction, abnormal heartbeat, and recovered myocardial infarction images). False-positive (FP) and false-negative (FN) are the misclassified other four classes (normal, myocardial infarction, abnormal heartbeat, and recovered myocardial infarction images) and COVID-19 ECG images, respectively.

The performance of deep CNNs was assessed using different evaluation metrics with 95% confidence intervals (CIs). Accordingly, CI for each evaluation metric was computed, as shown in equation (11):

$$r = z\sqrt{metric(1-metric)/N} \qquad (11)$$

where, N is the number of test samples, and $z$ is the level of significance that is 1.96 for 95% CI. In addition to the above metrics, the various classification networks were compared in terms of elapsed time per image, or the time it took each network to classify an input image, as shown in equation (12).

$$\Delta T = T2 - T1 \qquad (12)$$

In this equation, T1 is the starting time for a network to classify an image, I, and T2 is the end time when the network has classified the same image, I.

## 4. Results and Discussion

This section describes the performance of the different classification networks' performance on ECG trace image classification. The comparative performance of different CNNs for two-class, three-class, five-class classification schemes was shown in Table 4. It can be noted that for two and three class classification schemes DenseNet201 is outperforming while for the five-class classification InceptionV3 is showing the best performance.

Table 4: Comparison of the performances of the different CNN models for different classification schemes (Best result is presented as bold)

| Classification | Model | Result with 95% CI | | | | | Inference time |
|---|---|---|---|---|---|---|---|
| | | Overall | Weighted | | | | |
| | | Accuracy | Precision | Sensitivity | F1-score | Specificity | |
| 2 Class | MobieneV2 | 98.74 ± 0.52 | 98.74 ± 0.52 | 98.74 ± 0.52 | 98.73 ± 0.52 | 96.23 ± 0.88 | 0.18 |
| | Resnet18 | 98.62 ± 0.45 | 98.56± 0.35 | 98.6 ± 0.4 | 98.8 ± 0.44 | 96.21 ± 0.78 | 0.26 |
| | Resnet50 | 98.92 ± 0.48 | 98.93 ± 0.48 | 98.92 ± 0.48 | 98.91 ± 0.48 | 96.28 ± 0.87 | 0.44 |
| | Resnet101 | 99.01 ± 0.46 | 99.02 ± 0.46 | 99.01 ± 0.46 | 99 ± 0.46 | 96.59 ± 0.84 | 0.85 |
| | **Densenet201** | **99.1 ± 0.44** | **99.11 ± 0.43** | **99.1 ± 0.44** | **99.09 ± 0.44** | **96.9 ± 0.8** | **1.34** |
| | InceptionV3 | 98.78 ± 0.52 | 98.62 ± 0.54 | 100 ± 0 | 99.31 ± 0.38 | 95.2 ± 0.99 | 1.26 |
| 3 Class | MobieneV2 | 90.79 ± 1.34 | 91.26 ± 1.3 | 90.79 ± 1.34 | 90.76 ± 1.34 | 92.75 ± 1.2 | 0.22 |
| | Resnet18 | 92.81 ± 1.19 | 92.83 ± 1.19 | 92.81 ± 1.19 | 92.8 ± 1.19 | 94.44 ± 1.06 | 0.31 |
| | Resnet50 | 93.01 ± 1.18 | 93.09 ± 1.17 | 93.01 ± 1.18 | 93.01 ± 1.18 | 94.59 ± 1.05 | 0.48 |
| | Resnet101 | 93.02 ± 1.18 | 93.23 ± 1.16 | 93.01 ± 1.18 | 92.99 ± 1.18 | 94.54 ± 1.05 | 0.88 |
| | **Densenet201** | **97.36 ± 0.74** | **97.4 ± 0.74** | **97.36 ± 0.74** | **97.36 ± 0.74** | **97.93 ± 0.66** | **1.4** |
| | InceptionV3 | 96.89 ± 0.8 | 96.9 ± 0.8 | 96.9 ± 0.8 | 96.89 ± 0.8 | 97.6 ± 0.71 | 1.36 |
| 5 Class | MobieneV2 | 96.22 ± 0.88 | 96.29 ± 0.87 | 96.22 ± 0.88 | 96.2 ± 0.88 | 97.73 ± 0.69 | 0.25 |
| | Resnet18 | 95.34 ± 0.97 | 95.44 ± 0.96 | 95.34 ± 0.97 | 95.28 ± 0.98 | 97.02 ± 0.79 | 0.33 |
| | Resnet50 | 96.43 ± 0.86 | 96.43 ± 0.86 | 96.43 ± 0.86 | 96.4 ± 0.86 | 97.93 ± 0.66 | 0.52 |
| | Resnet101 | 97 ± 0.79 | 97.07 ± 0.78 | 97 ± 0.79 | 96.95 ± 0.79 | 97.97 ± 0.65 | 0.92 |
| | Densenet201 | 97.2 ± 0.76 | 97.2 ± 0.76 | 97.21 ± 0.76 | 97.2 ± 0.76 | 98.63 ± 0.54 | 1.61 |
| | **InceptionV3** | **97.83 ± 0.67** | **97.82 ± 0.67** | **97.83 ± 0.67** | **97.82 ± 0.67** | **98.86 ± 0.49** | **1.68** |

For two-class (Normal vs COVID-19) classification, overall test accuracy was 99.1% using Densenet201, while for three-class (Normal, COVID-19 and other cardiac abnormality) classification, it was 97.36% using Densenet201, and for five-class (normal, COVID-19, myocardial infarction, abnormal heartbeat, and recovered MI) classification, it was found to be 97.83% with InceptionV3. Figure 7 depicts the area under the curve (AUC) / receiver-operating characteristics (ROC) curve (also known as AUROC (area under the receiver operating characteristics) for various classification schemes, which is one of the most essential assessment metrics for determining the success of any classification model. In two-class and three-class classification, DenseNet201 shows better performance than the other techniques, as shown in Figure 7, where InceptionV3 outperforms the other algorithms for five-class classification.

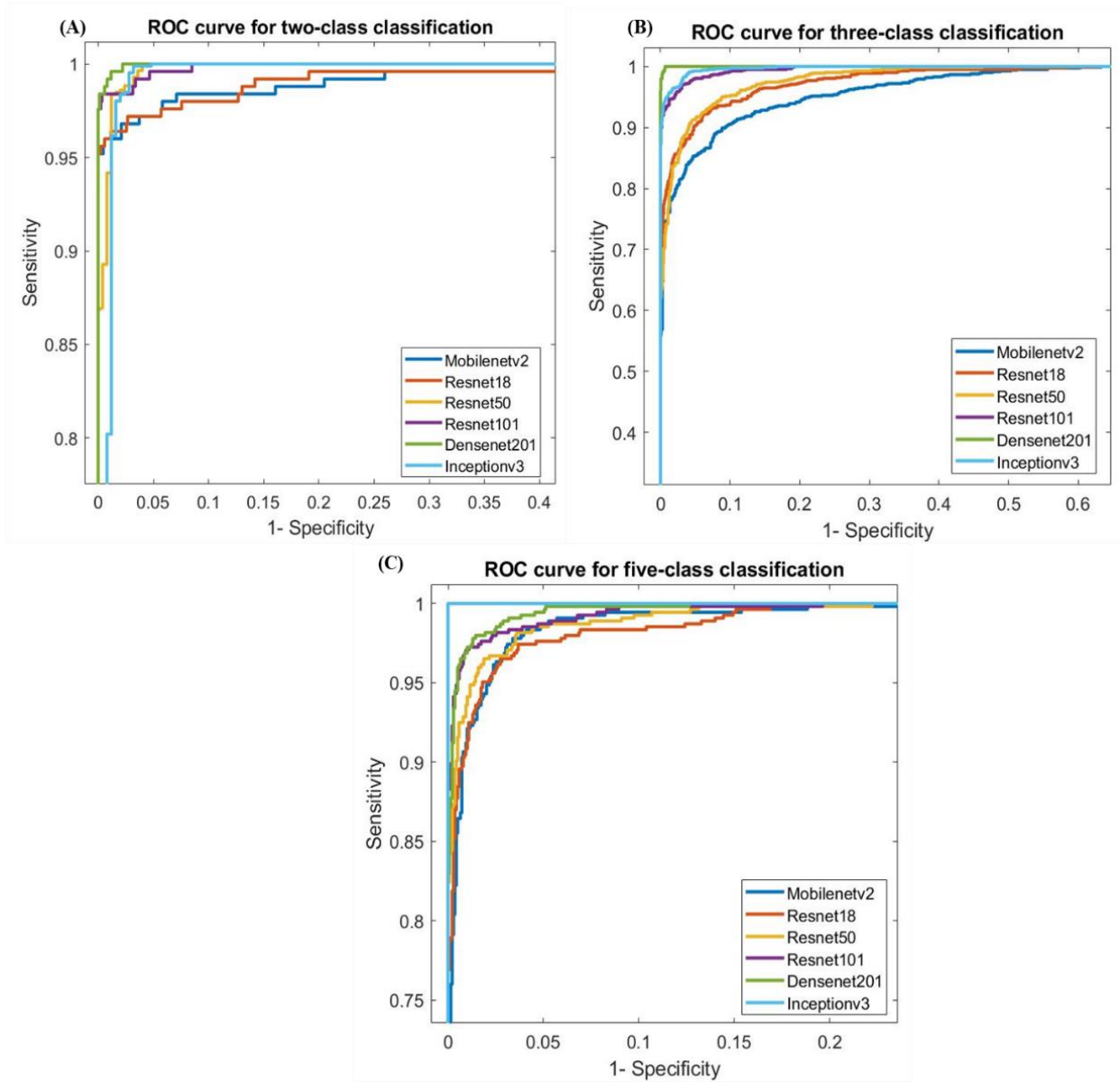

Figure 7: ROC curves for two-class, three-class, and five-class classifications for ECG images.

Figure 8 illustrates the confusion matrix for the outperforming model for ECG trace image classification schemes: two classes (Densenet201), three classes (Densenet201), and five classes (Inceptionv3). It is worth noting that with the top-performing network, 10 out of 250 COVID-19 ECG images were incorrectly categorized as normal for two-class classification, however, none of the COVID-19 ECG images was incorrectly categorized as normal or other classes for three-class or five-class classification. This is an outstanding performance from any computer-aided classifier, and this can significantly help in the fast diagnosis of COVID-19 by the clinicians immediately after acquiring the ECG images.

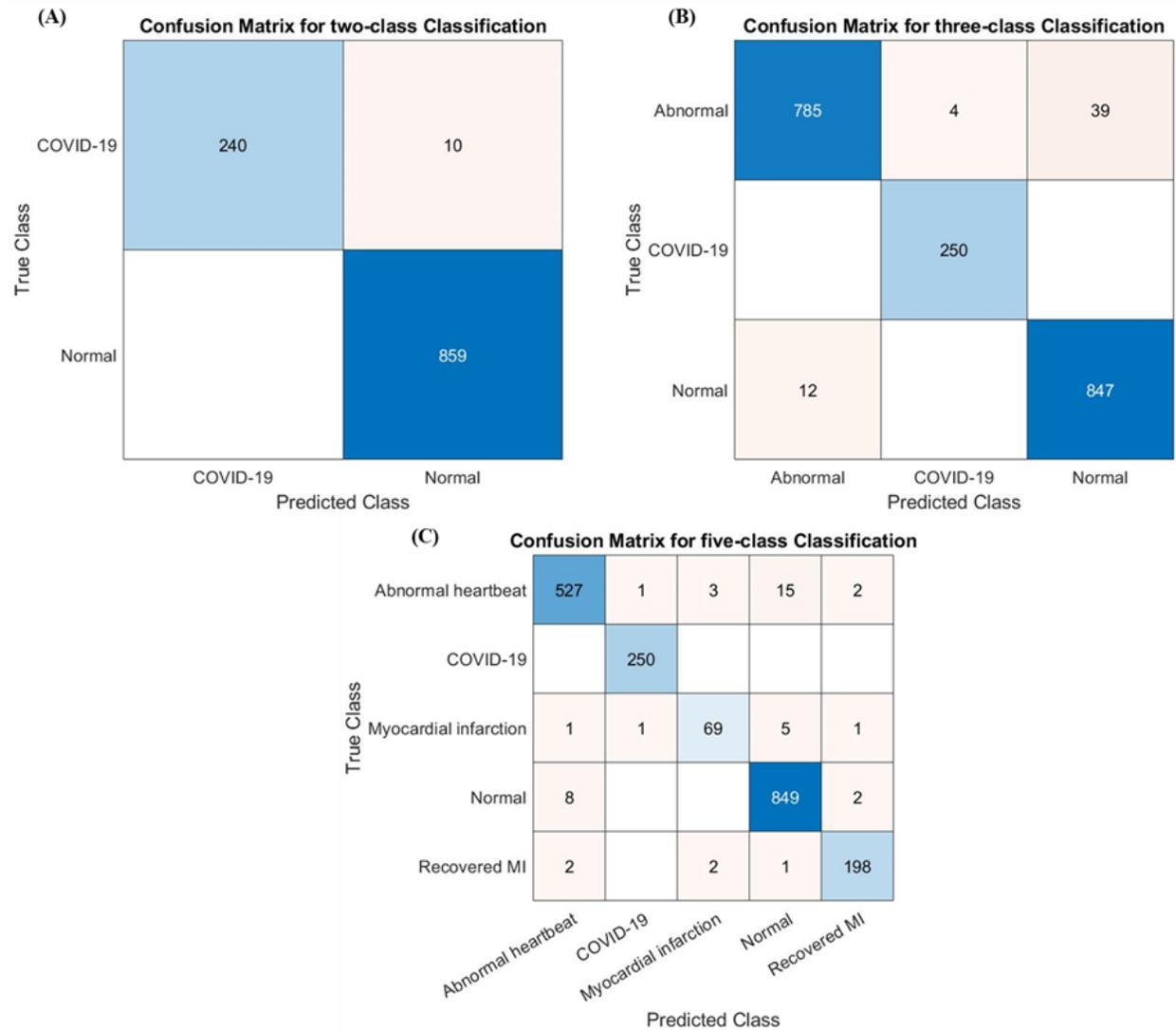

Figure 8: Confusion matrix for (A) Normal and COVID-19, (B) Normal, COVID-19, and Abnormal classification for Densenet201 model, and (C) Normal, COVID-19, Myocardial infarction, Abnormal heartbeat, and Recovered MI classification for Inceptionv3 model.

Figure 9 shows the comparison of accuracy versus the elapsed time per image for different CNN models for two-class, three-class, and five-class classification. While Densenet201 outperforms other networks for two-class, and three-class classification and Inceptionv3 outperform other networks for five-class classification, these are the slowest networks; however, these networks took approximately a second to take decision. For two-class and five-class classification, all network performances are comparable, where for three-class, Densenet201 outperforms other networks by 4-7%.

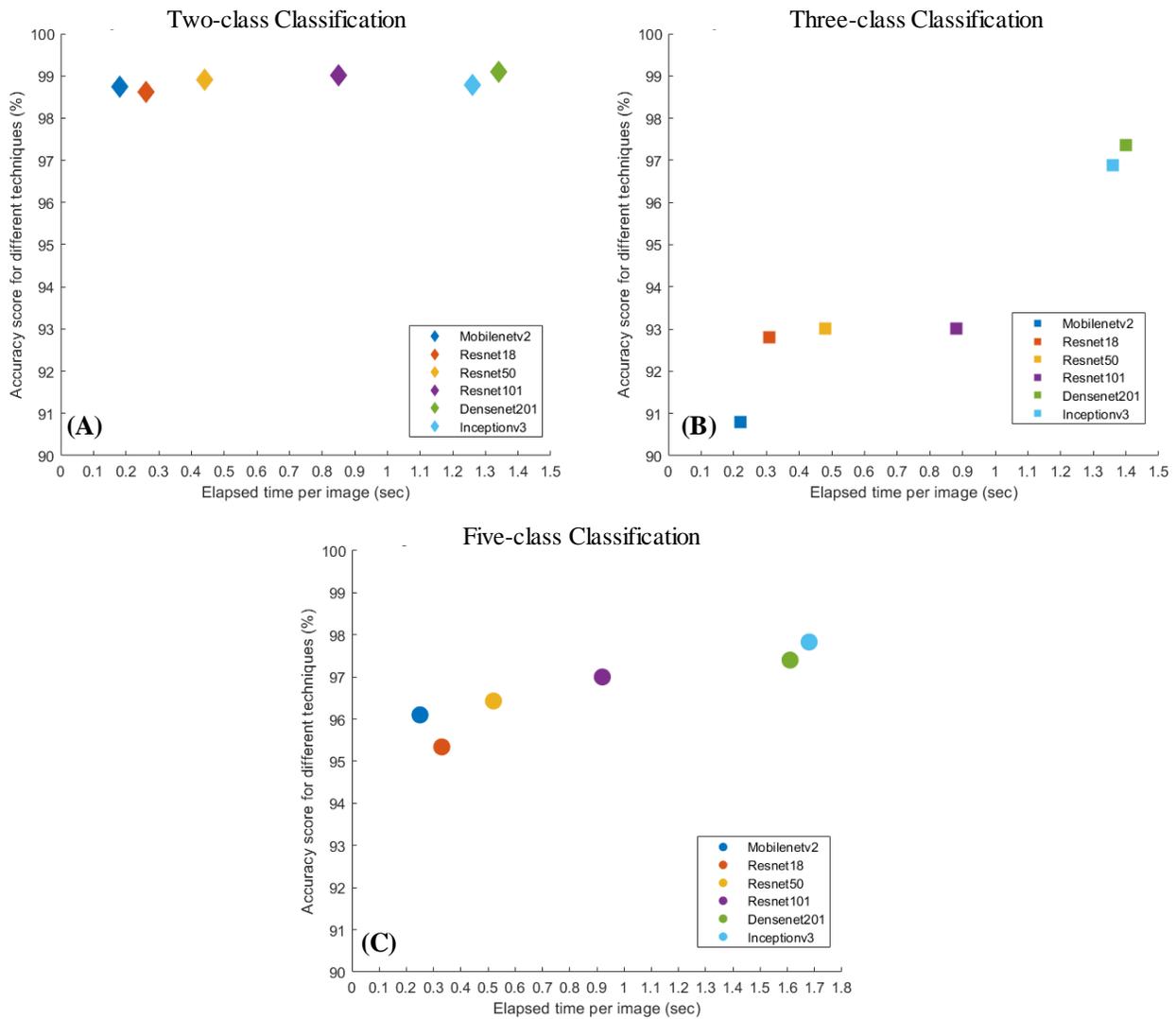

Figure 9: Accuracy vs inference time plot for two-class (A), three-class (B), and five-class (C) classifications.

Figure 10 shows the training and validation loss versus epochs for the three best-performing networks for two-class (DenseNet201), three-class (DenseNet201), and five-class (InceptionV3) classification. It can also be seen that the networks reach and stabilize with the lowest loss earlier after few epochs.

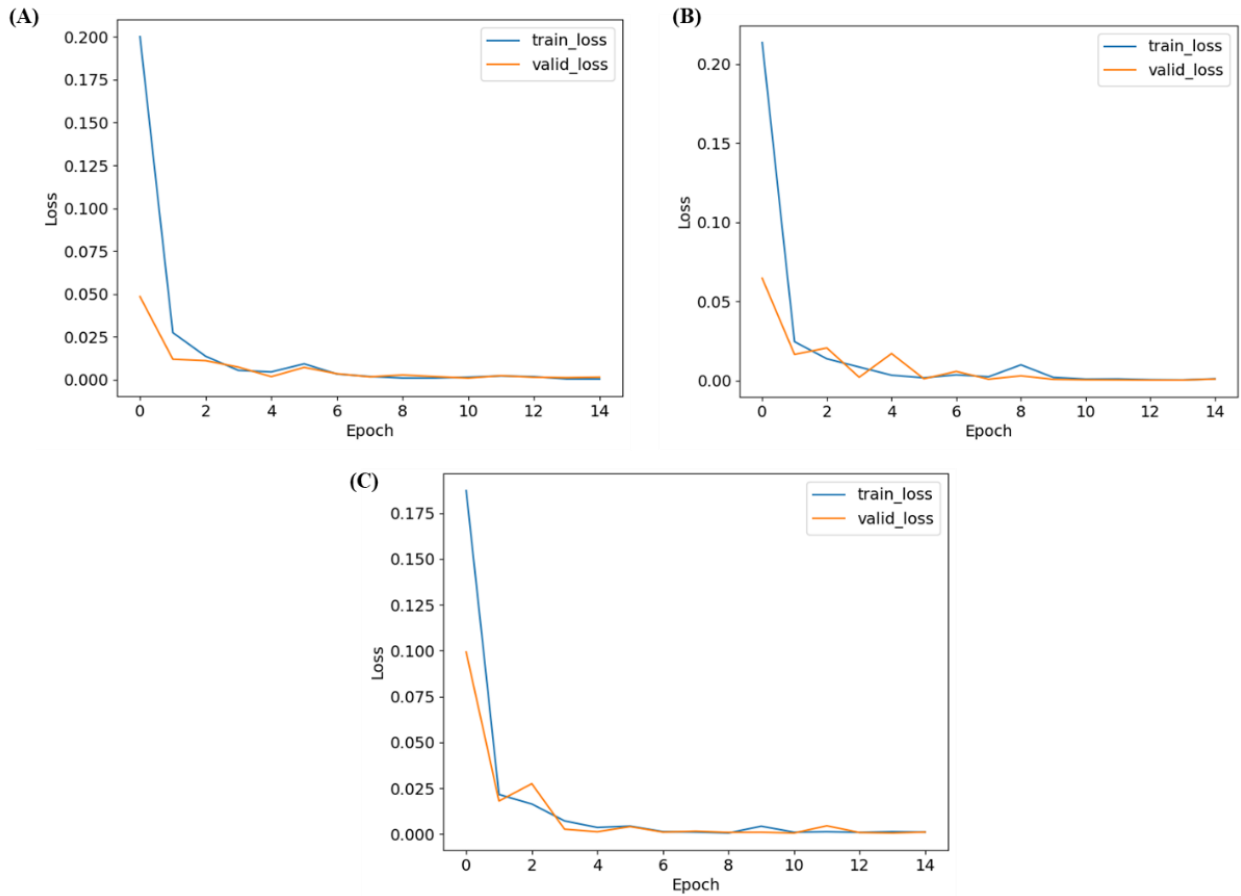

Figure 10: Training and Validation Losses versus Epoch for (A) two-class, (B) three-class, and (C) five-class classification.

As mentioned previously, it is critical to determine if the network is learning from the relevant area of the ECG trace images or from somewhere else and any non-related data for taking the decision. Heat maps based on the Score-CAM technique were created for distinct classes of the ECG trace images. Figure 11 depicts samples ECG trace images for 3-class classification as well as heat maps created using the best-performing DenseNet201 model. CNN learns from the regions where various waves change for various classes and the areas that are most important in determining abnormal ECG images in each of the ECG trace images. In Figure 11 (A-C), we can see that ST-segment and J-point elevation, and abnormal heartbeat occurred for COVID-19, myocardial infarction, and abnormal heartbeat. Reliability of how the network is taking decisions for classification is important to increase the confidence of the end-user in the AI performance. It is easily noticeable that the network learned from the area where ECG waves are changing compared to normal ECG images rather than the outside area of the ECG waves.

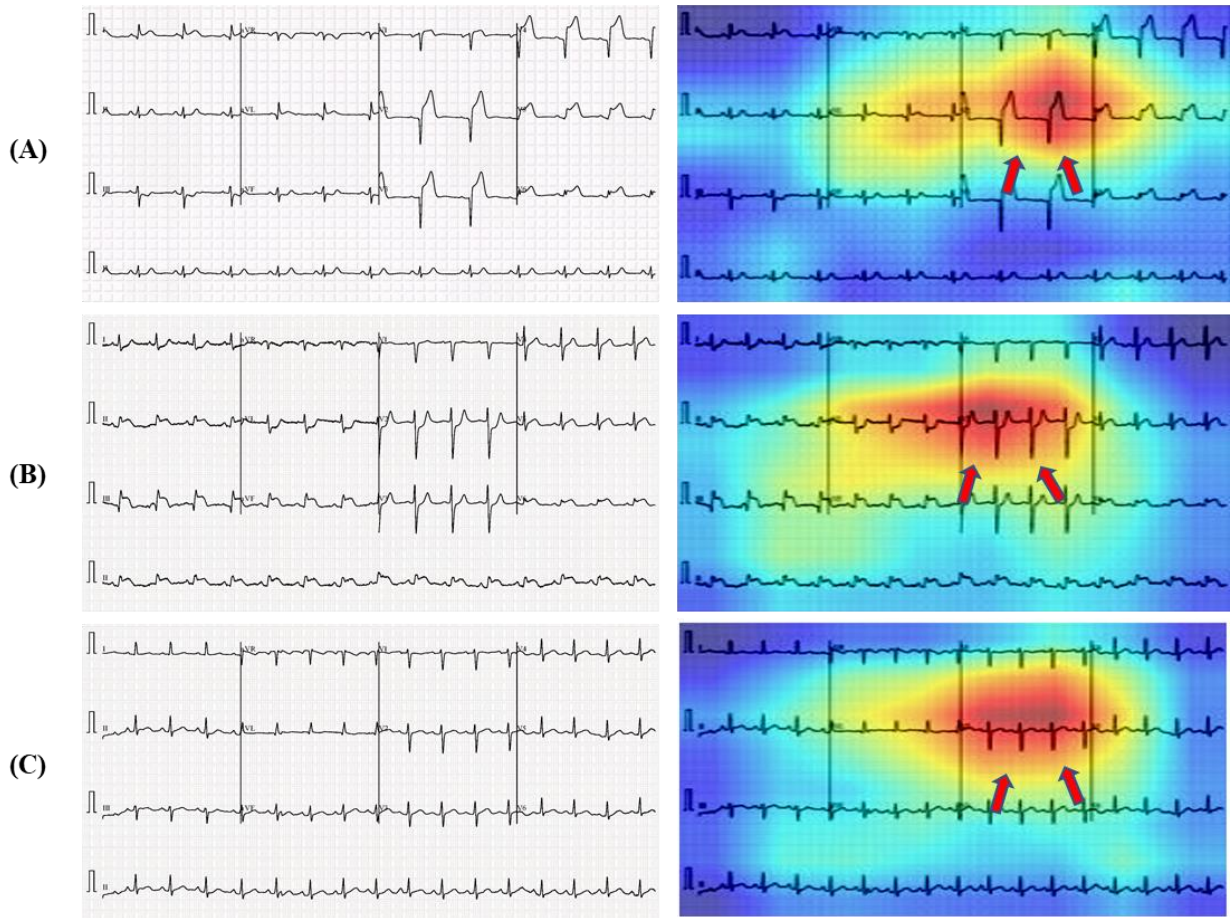

Figure 11: Score-CAM visualization of abnormal (COVID-19, myocardial infarction, and abnormal heartbeat) ECG images using the best performing model. The subtle signs are identified as key parts to detect abnormalities (highlighted in red and pointing via arrows).

## 5. Conclusion

This paper provides a deep Convolutional Neural Networks-based transfer learning strategy for the automated diagnosis of COVID-19 and other cardiac disorders using ECG trace images. The performance of the six different CNN models was evaluated for the classification of three different schemes: two-class classification (Normal and COVID-19), three-class classification (Normal, COVID-19, and Cardiac abnormality) and five-class classification (Normal, COVID-19, myocardial infarction (MI), Abnormal heartbeat (HB), and recovered MI). Densnet201 model outperforms other deep CNN models for two-class, and three-class classifications whereas Inceptionv3 outperform other networks for five-class classification. The best classification accuracy, precision, and recall for the two-class, and three-class classifications were found to be

99.1%, 99.11%, 99.1%, and 97.36%, 97.4%, 97.36%, respectively. For five-class classification, the best classification accuracy, precision, and recall were 97.82%, 97.83%, and 97.82%, respectively. The Score-CAM visualization output demonstrates that the important signal changes in the ECG trace contribute to the decision-making of the network. Automatic abnormality detection from ECG images has a very crucial application in computer-aided diagnosis for critical healthcare problems like this one. This state-of-the-art performance can be a very useful and fast diagnostic tool, which can save a significant number of people who died every year due to delayed or improper diagnosis.